\documentclass{jetpl}
\twocolumn

\lat

\usepackage{amsmath, amssymb}

\usepackage{graphicx}

\newcommand{\A}{\mathcal{A}}
\newcommand{\E}{\mathcal{E}}
\newcommand{\G}{\mathcal{G}}
\newcommand{\M}{\mathcal{M}}
\newcommand{\Hilb}{\mathbf{H}}
\renewcommand{\H}{\mathcal{H}}
\newcommand{\eps}{\varepsilon}

\newcommand{\dtau}{\partial_{\tau}}
\newcommand{\oD}[2]{\frac{\delta #1}{\delta #2}}

\title{Hamiltonian form and solitary waves of the spatial Dysthe equations}

\rtitle{Hamiltonian form of the spatial Dysthe equations}

\sodtitle{Hamiltonian form and solitary waves of the spatial Dysthe equations}

\author{F.\,Fedele$^+$\/\thanks{e-mail: fedele@gatech.edu}, D.\,Dutykh$^*$}

\rauthor{F.\,Fedele F., D.\,Dutykh D.}

\sodauthor{F.~Fedele, D.~Dutykh}

\address{$^+$School of Civil and Environmental Engineering, Georgia Institute of
Technology, Atlanta, USA\\~\\
$^*$LAMA, UMR 5127 CNRS, Universit\'e de Savoie, Campus Scientifique, 73376 Le Bourget-du-Lac Cedex, France}

\dates{16 October 2011}{*}

\abstract{The spatial Dysthe equations describe the envelope evolution of the free-surface and potential of gravity waves in deep waters. Their Hamiltonian structure and new invariants are unveiled by means of a gauge transformation to a new canonical form of the evolution equations. An accurate Fourier-type spectral scheme is used to solve for the wave dynamics and validate the new conservation laws, which are satisfied up to machine precision. Further, traveling waves are numerically constructed using the Petviashvili method. It is shown that their collision appears inelastic, suggesting the non-integrability of the Dysthe equations.}

\PACS{04.30.Nk, 47.10.Df, 47.35.Fg, 47.11.Kb}

\begin{document}

\maketitle

\section{INTRODUCTION}

Mathematical models used in physics and mechanics do not always possess a canonical Hamiltonian structure. Typically, the dynamics is governed by partial differential equations expressed in terms of physically-based variables, which are not usually canonical. A transformation to new variables is needed in order to unveil the desired structure explicitly. This is the case for the equations of motion for an ideal fluid: in the Eulerian description, they cannot be recast in a canonical form, whereas in a Lagrangian frame the Hamiltonian structure is revealed by Clebsch potentials. Moreover, multiple-scale perturbations of differential equations expressed in terms of non-canonical variables typically lead to approximate equations that do not maintain the fundamental conserved quantities, as the hydrostatic primitive equations on the sphere, where energy and angular momentum conservation are lost under the hydrostatic approximation. Clearly, if canonical variables can be identified, then the associated Hamiltonian structure provides a natural framework for making consistent approximations that preserve the fundamental dynamical properties of the original system, notably its conservation laws. For example, consider the equations that describe the irrotational flow of an ideal incompressible fluid of infinite depth with a free surface. Their Hamiltonian description was discovered by \cite{Zakharov1968} in terms of the free-surface elevation $\eta(x,t)$ and the velocity potential $\varphi(x,t) = \phi(x,z=\eta(x,t),t)$ evaluated at the free surface of the fluid. Variables $\eta(x,t)$ and $\varphi(x,t)$ are conjugated canonical variables with respect to the Hamiltonian $\H$ given by the total wave energy. By means of a third order expansion of $\H$ in the wave steepness, \cite{Zakharov1999} derived an integro-differential equation in terms of canonical conjugate Fourier amplitudes, which has no restrictions on the spectral bandwidth.

The modified Nonlinear Schr\"{o}dinger (NLS) equations derived by \cite{Dysthe1979} are also non-Hamiltonian. Using the method of multiple scales, he extended the deep-water cubic NLS\ equation for the time evolution of narrowband wave envelope $B$ of the carrier wave $\exp (ik_{0}x - i\omega_{0}t)$ and that of the potential $\phi$ of the wave-induced mean flow, to fourth order in steepness and bandwidth. Introducing dimensionless units, $t^{\prime} = \omega_{0}t$, $x^{\prime} = k_{0}x$, $B^{\prime} = k_{0}B$, and dropping the primes, $\phi$ can be easily found by means of the Fourier-transform and a single equation for $B$ can be derived as well:
\begin{multline}\label{eq:envB}
B_t + \frac{1}{2}B_{x}+\frac{1}{8}iB_{xx}-\frac{1}{16}B_{xxx}+\frac{i}{2}%
\left\vert B\right\vert ^{2}B+\frac{3}{2}|B|B_{x} \\
+ \beta B^{2}B_{x}^{\ast} + \frac{1}{2}iB\Hilb(|B|^2)_{x} = 0,
\end{multline}%
where\ $\beta =1/4$, the subscripts $B_{t}=\partial _{t}B$ and $B_{x}=\partial _{x}B$ denote partial derivatives with respect to $x$ and $t$, respectively, $\Hilb(f)$ is the Hilber Transform of a function $f(x)$, and $B^{\ast}$ denotes complex conjugation (see also \cite{Janssen1983}). The form of the equation for the envelope $A$ of the wave potential is similar to (\ref{eq:envB}), but the term $\beta B^{2}B_{x}^{\ast}$ becomes $-\beta A^{2}A_{x}^{\ast}$ \cite{Hogan1985}. Recently, the associated canonical form (which do not contain the $\beta$ term)\ has been derived by \cite{Gramstad2011} starting from the Hamiltonian Zakharov equation with the Krasitskii kernel \cite{Krasitskii1994}. We also point out that \cite{Zakharov2010}, starting from a conformal-mapping formulation of the Euler equations derived another version of the temporal Dysthe equation, which is similar to (\ref{eq:envB}) but also non-Hamiltonian.

On the other hand, to model wave propagation in wave basins a change to a coordinate system moving at the group velocity can be used by introducing the dimensionless variables%
\[
  B=\eps u,\qquad \tau =\eps (2x-t),\qquad \xi =\eps^{2}x,
\]
with $\eps = k_{0}a$ being the wave steepness of the carrier wave and $a$ the associated amplitude \cite{Lo1985}. As such, the temporal Dysthe equation (\ref{eq:envB}) transforms, up to the fourth order in $\eps$, to
\begin{equation}\label{eq:SK}
u_{\xi }+iu_{\tau \tau }+i\left\vert u\right\vert ^{2}u+8\eps \left\vert
u\right\vert^2 u_{\tau }+2\eps u^{2}u_{\tau }^{\ast } + 
2\eps iu\Hilb(\left\vert u\right\vert ^{2})_{\tau } = 0,
\end{equation}%
hereafter referred to as the spatial Dysthe for the wave envelope $u$. On the other hand, the associated envelope $A=\eps v$ of the wave potential satisfies
\begin{equation}\label{eq:PO}
  v_{\xi }+iv_{\tau \tau }+i|v|^2v + 8\eps|v|^2 v_{\tau} + 
  2\eps iv\Hilb(|v|^2)_\tau = 0.
\end{equation}
Both (\ref{eq:SK}) and (\ref{eq:PO}) can also be derived directly from the Zakharov equation \cite{Kit2002}. 

In this paper, we will unveil the hidden canonical structure of the spatial Dysthe equations (\ref{eq:SK}), (\ref{eq:PO}). In particular, we introduce a gauge transformation that yields a canonical form of equation (\ref{eq:SK}) for the wave envelope $u$, and new invariants for it. As a corollary, we will also show that equation (\ref{eq:PO}) for the wave potential envelope $v$ is already Hamiltonian. Then, the Petviashvili method is exploited to compute numerically ground states and traveling waves (\cite{Petviashvili1976}; see also \cite{Lakoba2007, Yang2010}). Their dynamics is numerically investigated up to machine precision by means of a highly accurate pseudo spectral scheme in order to provide new insights on the integrability of the Dysthe equations.

\section{CANONICAL FORM}

Hereafter, we will consider the generic nonlinear equation
\begin{multline}\label{eq:we}
  u_{\xi } = -ia u_{\tau\tau} - ih\left\vert u\right\vert ^{2}u 
  - c\eps\left\vert u\right\vert ^{2}u_{\tau } \\
  -\eps eu^{2}u_{\tau }^{\ast} - fi\eps u \Hilb(\left\vert u\right\vert ^{2})_{\tau},
\end{multline}
with $(a,h,c,e,f)$ as a quintuplet of arbitrary real coefficients. In particular, the spatial Dysthe for the wave and potential envelopes follow from (\ref{eq:we}) with parameters $\left(1,1,8,2,2\right) $ and $\left(1,1,8,0,2\right)$ respectively. The wave action 
\begin{equation}\label{eq:Ac}
  \A = \int \left\vert u\right\vert ^{2}d\tau
\end{equation}
is conserved by (\ref{eq:we}), but up to date no other conservation laws are known for $u$.

Drawing from \cite{Colin2006} (see also \cite{Wyller1998}), the invariance of $\A$ suggests the following variable change via the gauge transformation%
\begin{equation}\label{eq:9}
 w = \G(u) = u\exp(ik\psi ),
\end{equation}
where $k$ is a free parameter, and the 'stream function' $\psi $ is defined as $\dtau\psi = \psi_\tau = |u|^2$. Note that $\left\vert u\right\vert^{2}=\left\vert w\right\vert^{2}$ and the wave action $\A$ is preserved in the transformation. The spatial evolution equation for $w$ follows from (\ref{eq:9}) as
\begin{multline}\label{eq:10}
  w_{\xi} = -iaw_{\tau\tau} - ih|w|^2w + \\ 
  ik\eps^2\frac{c-3e-2ak}{2} |w|^4w - (c+2ak)\eps |w|w_{\tau} \\ 
  - (e+2ak)\eps w^2w_{\tau}^{\ast} - if\eps w\Hilb(|w|^2)_{\tau}.
\end{multline}
If the free parameter $k^{\ast}$ is chosen as $k^{\ast} = -\frac{e}{2a}\eps$, equation (\ref{eq:10}) simplifies to
\begin{multline}\label{eq:11}
 w_{\xi } = -iaw_{\tau \tau }-ih\left\vert w\right\vert ^{2}w-i\frac{ce-2e^{2}}{4a}\eps^{2}|w|^{4}w \\ -(c-e)\eps \left\vert
 w\right\vert ^{2}w_{\tau }-if\eps w\Hilb(\left\vert w\right\vert^{2})_{\tau},
\end{multline}
which admits the Hamiltonian structure
\begin{equation}
 \left( 
 \begin{array}{c}
  w_{\xi } \\ 
  w_{\xi }^{\ast }
 \end{array}
 \right) = i\left( 
 \begin{array}{cc}
  0 & 1 \\ 
  -1 & 0%
 \end{array}
 \right) \left( 
  \begin{array}{c}
   \oD{\H_{w}}{w} \\ 
   \oD{\H_{w}}{w^{\ast}}
 \end{array}
 \right) ,
\end{equation}
where the Hamiltonian is given by
\begin{multline}\label{eq:12}
  \H_{w} = \int \Bigl( a\left\vert w_{\tau }\right\vert ^{2}-\frac{h}{2}
 \left\vert w\right\vert ^{4}-\frac{ce-2e^{2}}{12a}\eps ^{2}\left\vert
 w\right\vert^{6} \\
 -i\frac{c-e}{4}\eps \left\vert w\right\vert
 ^{2}(w_{\tau }^{\ast }w - w_{\tau }w^{\ast }) - \frac{f}{2}\eps \left\vert
 w\right\vert ^{2}\Hilb(\left\vert w\right\vert ^{2})_{\tau }\Bigr) d\tau.
\end{multline}
Here, $\H_{w}$ is also an invariant together with the momentum 
\begin{equation}\label{eq:M}
  \M_{w} = \int i(w_{\tau }^{\ast }w-w_{\tau }w^{\ast })d\tau .
\end{equation}%
Thus, from (\ref{eq:9})
\begin{multline}\label{eq:13}
 \E(u) = \int\Bigl( a|u_{\tau}|^2 - \frac{h}{2}|u|^4 + \frac{ce+e^2}{6a}\eps^2|u|^6 \\
 -i\frac{c+e}{4}\eps \left\vert u\right\vert ^{2}(u_{\tau }^{\ast}u - u_{\tau}u^{\ast })-\frac{f}{2}\eps \left\vert u\right\vert^{2}\Hilb(\left\vert u\right\vert ^{2})_{\tau }\Bigr) d\tau,
\end{multline}
\[
 \M(u) = \int \left[ i(u_{\tau }^{\ast }u-u_{\tau }u^{\ast })-\frac{e}{a}\eps\left\vert u\right\vert ^{4}\right] d\tau,
\]
are both invariants of the spatial Dysthe equation (\ref{eq:we}), but $\E$ is not the associated Hamiltonian. Note the appearance of terms of $O(\eps^2)$ in the invariants $\E$ and $\H_w$. They both vanish if $e$ is null; as a result $u = w$ and $\E(u) = \H_w$ becomes the Hamiltonian for $u$. As a consequence, the spatial Dysthe (\ref{eq:envB}) for the wave envelope $B$ is non Hamiltonian since $e = 2$, whereas the wave potential $v$ is canonical (cf. eq. (\ref{eq:PO})). 

In the next section we use a highly accurate Fourier-type pseudo-spectral method to solve for the envelope dynamics and validate the invariance of the Hamiltonian (\ref{eq:12}) of $w$ and the new invariants (\ref{eq:13}) of $u$. In our numerical investigations, it is found that they are conserved up to machine precision.

\section{GROUND STATES AND TRAVELING WAVES}

Insights into the underlying dynamics of the Dysthe equations are to be gained if we construct some special families of solutions in the form of ground states and traveling waves, often just called solitons or solitary waves. Hereafter, we do so for the Dysthe equation (\ref{eq:we}) for the envelope $u$, but the associated Hamiltonian form (\ref{eq:11}) in $w$ can be treated in a similar way. However, owing to the gauge transformation (\ref{eq:9}) the envelope $|w| = |u|$. We point out that analytical solutions of (\ref{eq:we}) are available in terms of multiple-scale perturbations of the NLS equation using, for example, direct soliton perturbation theory (see, for example, \cite{Akylas1989, Yang2010}). Consequently, we construct numerical solutions of ground states and traveling waves of the form $u(\xi, \tau) = v(\tau - s\xi)e^{-i\mu\xi}$ using the Petviashvili method (\cite{Petviashvili1976}, see also \cite{Lakoba2007}), where $\mu$ and $s$ are generic parameters and the function $v(\cdot)$ is in general complex. This numerical approach has been successfully applied by \cite{Zakharov2010} to compute ground states of their version of the temporal Dysthe equation.

\begin{figure}
  \centering
  \includegraphics[width=0.49\textwidth]{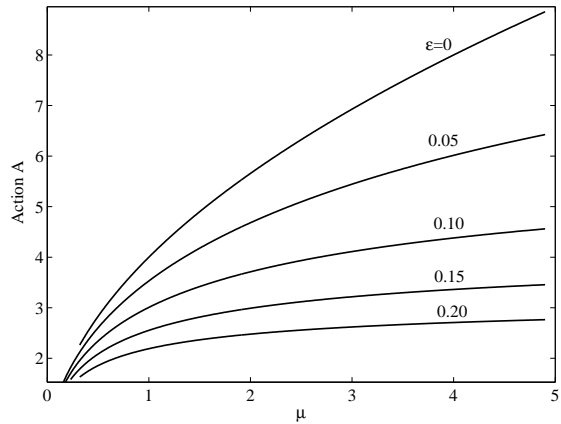}
  \caption{\textbf{Figure 1.} Action $\A$ of the ground state of the spatial Dysthe equation (\ref{eq:SK}) for $u$ as function of $\mu$, for different values of the steepness $\eps$.}
  \label{fig:fig1}
\end{figure}

\begin{figure}
  \centering
  \includegraphics[width=0.49\textwidth]{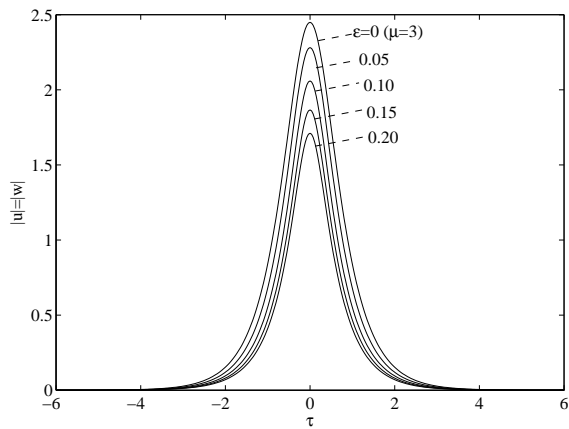}
  \caption{\textbf{Figure 2.} Envelopes $|u|$ of ground states of the spatial Dysthe equation (\ref{eq:SK}) for different values of the steepness $\eps$ ($\mu = 3$). Note that $|u|=|w|$ due to the gauge invariance (\ref{eq:9}).}
  \label{fig:fig2}
\end{figure}

As an application, consider the non-Hamiltonian Dysthe (\ref{eq:SK}), particular case of (\ref{eq:we}) with parameters $(1,1,8,2,2)$. Figure \ref{fig:fig1} shows the action $\A$ of the ground state ($s=0$) for $u$ (and so $w$ due to the gauge invariance of (\ref{eq:9})) as function of $\mu$, computed for different values of the steepness $\eps$, and Figure \ref{fig:fig2} reports the associated envelopes $|u| = |w|$ for $\mu = 3$. As one can see, as $\eps$ increases they tend to reduce in size in agreement with the asymptotic analysis carried out by \cite{Akylas1989} in the limit of $\eps\to 0$.

\begin{figure}
  \centering
  \includegraphics[width=0.49\textwidth]{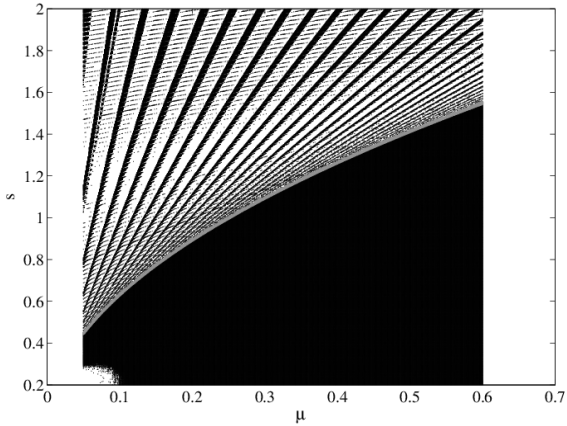}
  \caption{\textbf{Figure 3.} Numerical basin of attraction of the Petviashvili scheme in the phase space $(s,  \mu)$ for $\eps = 0.15$. Solitary waves (localized traveling waves) occur below the gray boundary curve $\Gamma$, which separates the region of periodic waves.}
  \label{fig:fig3}
\end{figure}

\begin{figure}
  \centering
  \includegraphics[width=0.49\textwidth]{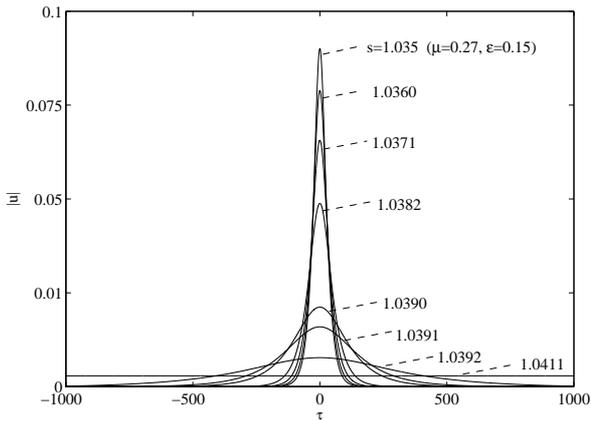}
  \caption{\textbf{Figure 4.} Localized traveling waves bifurcating from the boundary $\Gamma$ of Figure \ref{fig:fig3}. Change in shape of the envelope $|u|$ as $s$ varies while keeping $\mu = 0.27$. In particular, with reference to Figure \ref{fig:fig3}, the point $(s, \mu = 0.27)$ reaches the boundary $\Gamma$ from below.}
  \label{fig:fig4}
\end{figure}

Furthermore, Figure \ref{fig:fig3} illustrates a typical basin of attraction of the Petviashvili scheme in the phase space $(s, \mu)$ for $\eps = 0.15$. Each black dot corresponds to a well converged solution up to machine precision, whereas white spots are associated to either divergent or converged-to-zero solutions. In particular, we noted that the numerical scheme converged to localized traveling waves below the gray boundary curve $\Gamma$ shown in Figure \ref{fig:fig3}. On the other hand, the convergence to simple periodic waves occurred for points above $\Gamma$. This curve agrees with the analytical form derived from the exact nonlinear dispersion of periodic waves, which follows from (\ref{eq:we}) as $s = (c-e)A^{2} + 2\sqrt{a(\mu - hA^2)}$, where $A$ is the wave amplitude. Thus, localized traveling waves bifurcate from $\Gamma$. This is clearly illustrated in Figure \ref{fig:fig4}, which reports the change in shape of the envelope $|u|$ as $s$ varies while keeping $\mu = 0.27$. In particular, with reference to Figure \ref{fig:fig3}, as the point $(s, \mu = 0.27)$ reaches the boundary $\Gamma$ from below, the soliton envelope tends to flatten to that of a periodic wave. Similar results also hold for the Hamiltonian $w$ due to the gauge transformation (\ref{eq:9}).

\begin{figure}
  \centering
  \includegraphics[width=0.49\textwidth]{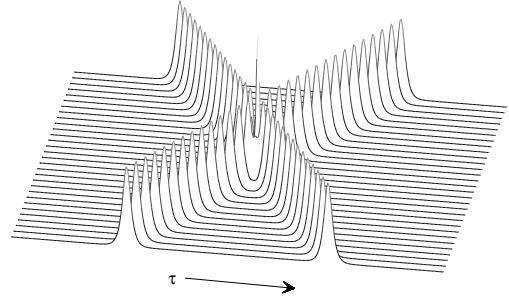}
  \caption{\textbf{Figure 5.} Elastic collision of two NLS solitary waves traveling at the same speed $s=2$, for $\mu = 2$.}
  \label{fig:fig5a}
\end{figure}

\begin{figure}
  \centering
  \includegraphics[width=0.49\textwidth]{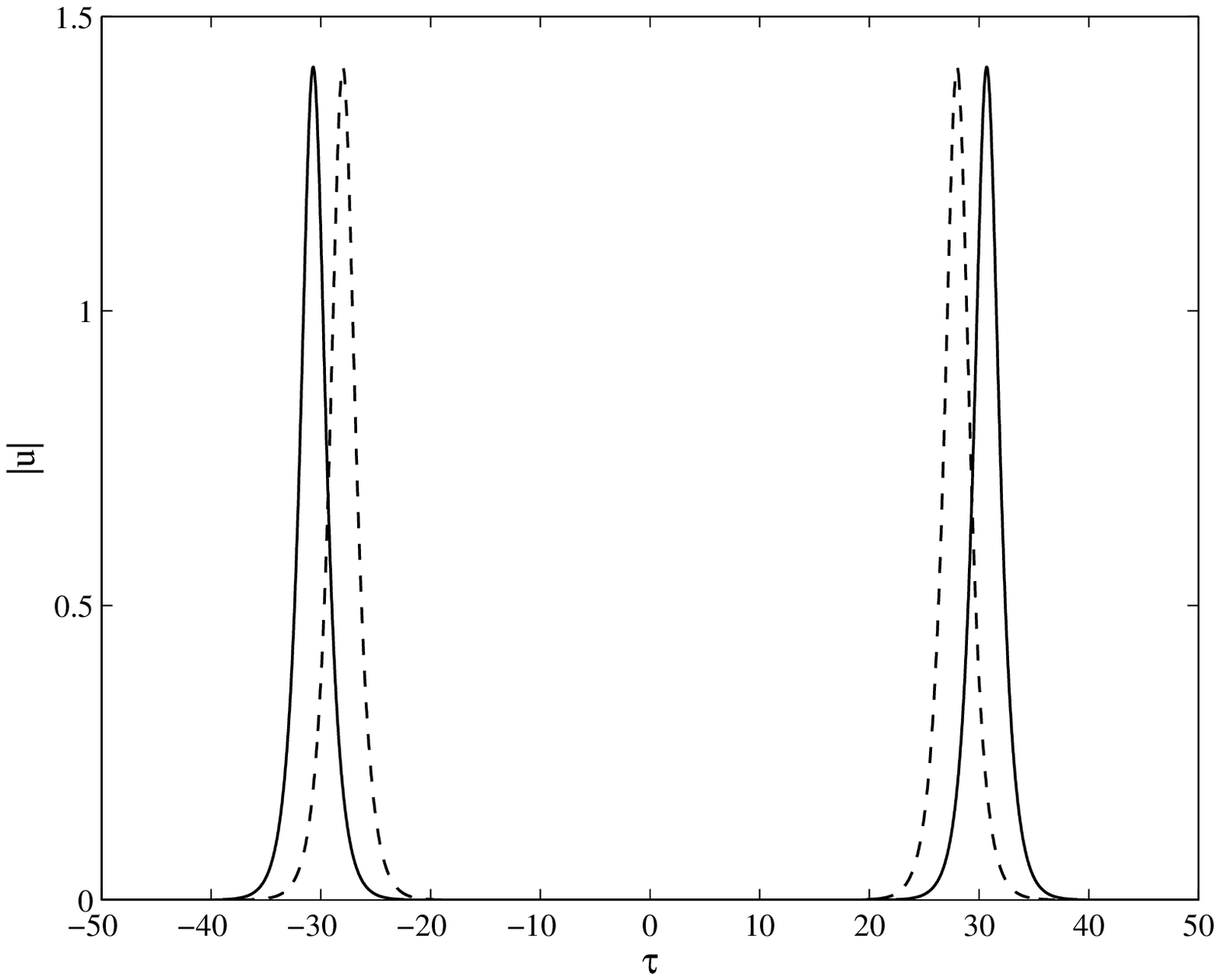}
  \caption{\textbf{Figure 6.} Initial ($--$) and final (---) shapes of one of the two solitons after the elastic collision of two NLS solitary waves traveling at the same speed $s=2$, for $\mu = 2$.}
  \label{fig:fig5b}
\end{figure}

\begin{figure}
  \centering
  \includegraphics[width=0.49\textwidth]{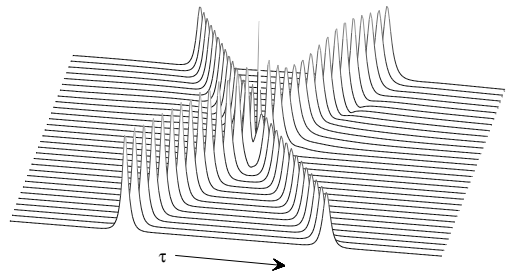}
  \caption{\textbf{Figure 7.} Inelastic collision of two Dysthe solitary waves traveling at the same speed $s=2$, for $\mu = 2$.}
  \label{fig:fig6a}
\end{figure}

\begin{figure}
  \centering
  \includegraphics[width=0.49\textwidth]{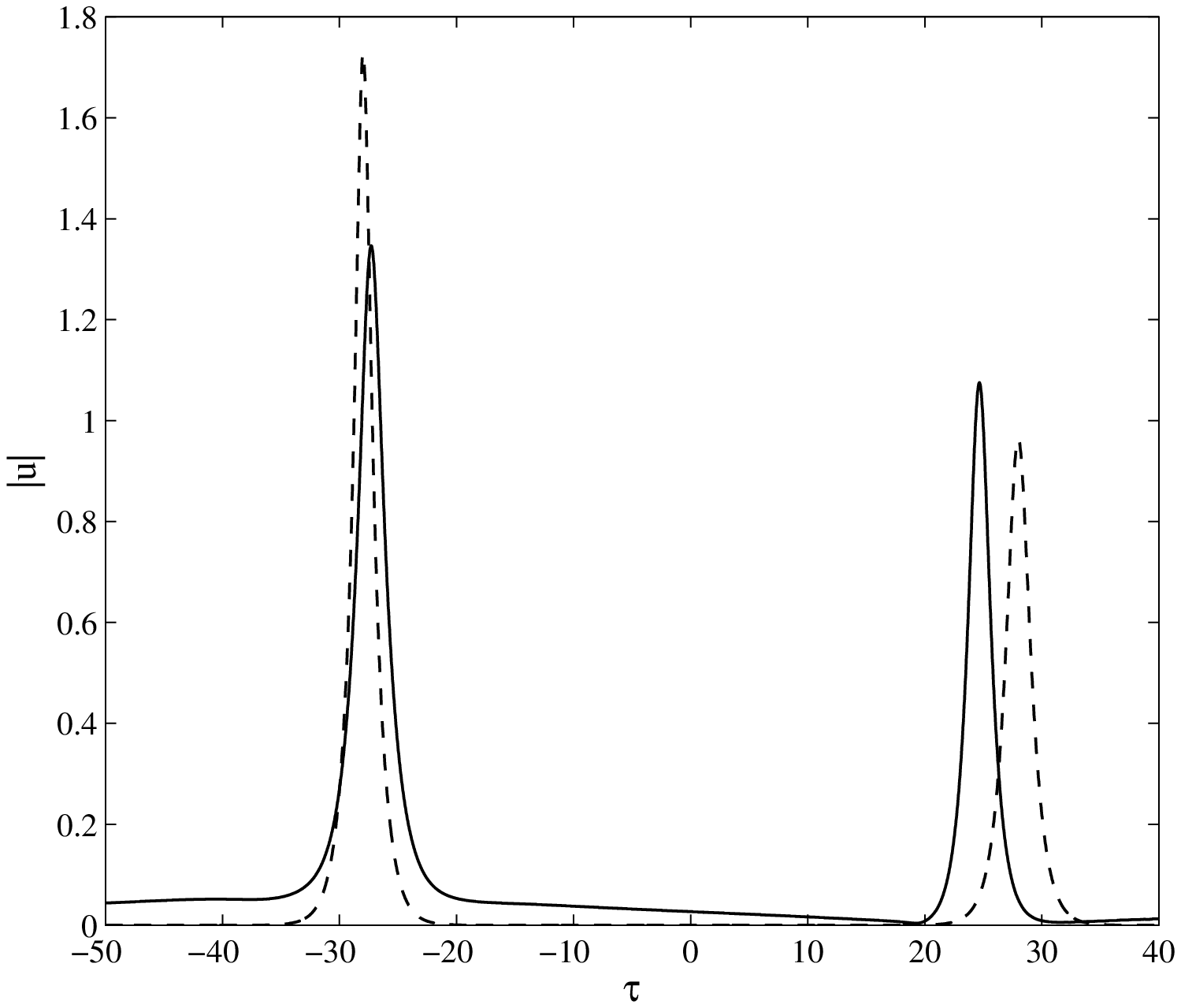}
  \caption{\textbf{Figure 8.} Initial ($--$) and final (---) shapes of one of the two solitons after an inelastic collision of two Dysthe solitary waves traveling at the same speed $s=2$, for $\mu = 2$.}
  \label{fig:fig6b}
\end{figure}

\begin{figure}
  \centering
  \includegraphics[width=0.49\textwidth]{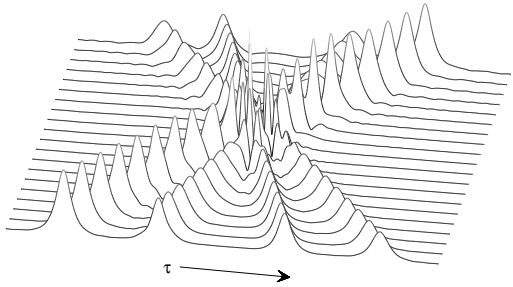}
  \caption{\textbf{Figure 9.} Inelastic collision of four Dysthe solitary waves.}
  \label{fig:fig7}
\end{figure}

Hereafter, we investigate the collision of traveling waves by means of a highly accurate Fourier-type pseudo-spectral method (see, for example, \cite{Clamond2001}). As an application, consider the interaction between two solitary waves of the NLS equation traveling in opposing directions with the same speed $s = 2$, for $\mu = 2$ ($\eps = 0$). The shape of the two solitons is identical because of the NLS reflection symmetry, i.e. $u(\xi, \tau) = (\xi, -\tau)$. The plot of Figure \ref{fig:fig5a} shows that the two solitons emerge out of the collision with the same shape, but a phase shift. The interaction is elastic as it should be since the NLS equation is integrable. This is clearly seen from the plot of Figure \ref{fig:fig5b}, which reports the initial and final shapes of the two solitary waves. For the same parameters, Figure \ref{fig:fig6a} reports the interaction of the associated solitary waves of the Dysthe equation (\ref{eq:SK}) for $\eps = 0.15$. The reflection symmetry is lost and the two solitons have different shape and amplitude. The interaction is clearly inelastic, since after the collision radiation is shed and the initial and final soliton shapes are different as seen in Figure \ref{fig:fig6b}. The interaction of four solitons has similar inelastic characteristics as shown in Figure \ref{fig:fig7}. This suggests the non-integrability of the Dysthe equation (\ref{eq:SK}). Similar dynamics is also observed for the associated Hamiltonian form (\ref{eq:11}).

\section{CONCLUSIONS}

A canonical variable for the spatial Dysthe equation for the wave envelope $u$ has been identified by means of the gauge transformation (\ref{eq:9}), and the hidden Hamiltonian structure is unveiled. Moreover, the gauge invariance yields two new invariants for the noncanonical $u$. It is also found that the envelope $v$ of the associated wave potential is canonical.

Further, the existence of solitary waves that bifurcate from periodic waves has been investigated numerically by means of a highly accurate Petviashvili scheme. In particular, ground state solutions are in agreement with the asymptotic analysis of \cite{Akylas1989}. Finally, the envelope dynamics has been investigated by means of  highly accurate Fourier-type pseudo-spectral method up to machine precision. It is found that solitary waves interact inelastically, suggesting the non-integrability of both the Hamiltonian and non-Hamiltonian version of the spatial Dysthe equations.

\bibliographystyle{plain}
\bibliography{biblio}

\end{document}